\documentclass[floats,twocolumn,prx,superscriptaddress]{revtex4-2}
\usepackage{graphics}
\usepackage{amsmath}
\usepackage{graphicx}
\usepackage{amsfonts}
\usepackage{amssymb}
\usepackage{eurosym}
\usepackage{graphics}
\usepackage{float}
\usepackage{epsfig}
\usepackage{latexsym}
\usepackage{theorem}
\usepackage{bm}
\usepackage{ulem}
\usepackage{color}
\usepackage{graphicx}
\usepackage{subcaption} 

\begin{document}

\title{Optimizing Epsilon Security Parameters in QKD}
\author{Alexander G. Mountogiannakis}
\affiliation{Department of Computer Science, University of York, York YO10 5GH, United Kingdom}
\affiliation{nodeQ, 71-75 Shelton Street, Covent Garden, London WC2H 9JQ, United Kingdom}
\author{Stefano Pirandola}
\affiliation{Department of Computer Science, University of York, York YO10 5GH, United Kingdom}

\begin{abstract}
    We investigate the optimization of $\varepsilon$-security parameters in quantum key distribution (QKD), aiming to improve the achievable secure key rate under a fixed overall composable security level. For this purpose, we employ a continuous genetic algorithm (CGA) to optimize the $\varepsilon$-security components of two representative protocols: the homodyne protocol from the continuous-variable (CV) family and the BB84 protocol from the discrete-variable (DV) family. We detail the CGA configuration, summarize the derivation of the composable key rate, and emphasize the role of the $\varepsilon$-parameters in both protocols. We then compare key rates obtained with optimized $\varepsilon$-values against those derived from standard and randomized choices. Our results demonstrate substantial key rate improvements at high security levels, where the key rate typically vanishes, and uncover positive-rate regimes that are inaccessible without optimization.
\end{abstract}

\maketitle

\section{Introduction}
Quantum key distribution (QKD) attempts to enable two parties to securely generate and share a secret cryptographic key by exploiting the principles of quantum mechanics \cite{PirandolaEtAl2020,Gisin,Feihu,scarani2009, CV-QKD-vlad, CV-MDI-QKD-review,Renner,TomamichelThesis} while being affected by a fundamental rate limitation~\cite{SKC,PLOB}. As the security of QKD is based on physics and not computational complexity, it is often described as providing ``unconditional security''~\cite{Mayers2001}. The original QKD paper was presented in 1984~\cite{bb84} and, since then, the field has grown substantially, with numerous theoretical advancements, experimental implementations, and the development of multiple security proofs against various types of attacks. The field is generally divided into discrete-variable (DV)~\cite{Ekert,ShorPreskill2000,BB84_general1,BB84_general2,BB84_general3,Lim,PortmannAndRenner2014,Yin, composable} and continuous-variable (CV) QKD~\cite{Ralph99, Hillery00, Cerf01, Ralph02, Fred02, Fred06, pirandola2008,
Lev2010, Vlad14, Vlad15, Lev2015, Vlad16, Zhang2020, Lev2017, pirandola2021,
pirandola2021b, Homodyne, Heterodyne, CV-MDI, ImprovedRates,
LeverrierEtAl2010, leverrier2010}.

Initially, QKD relied on the notion of the asymptotic key rate, which examines the protocol behavior, when an infinite number of quantum states are transmitted \cite{ShorPreskill2000}. As the practical demonstration of QKD protocols unfolded, security solely in the asymptotic regime became obsolete. Finite-size effects, such as the number of generated states or the deviation between estimated and actual values, started to be incorporated into security proofs. However, these proofs remained incomplete, as practical implementations involve hidden imperfections in various components of the protocols that reduce their security. To mitigate the effects of these imperfections, the concept of composability was introduced to QKD \cite{Renner}.

Composability is associated with building systems, whose security is preserved, when protocols are composed with other protocols or used as components in larger applications \cite{composable}. In the context of quantum cryptographic systems, a protocol is $\varepsilon$-secure, when \cite{Renner}
\begin{equation}
    \mathcal{D} = (\rho_{ABE}, \sigma_\text{AB} \otimes \rho_E) \leq \varepsilon,
\end{equation}
where $\mathcal{D}$ is the trace distance, $\rho_{ABE}$ is the joint output state of Alice, Bob, and Eve, and $\sigma_\text{AB} \otimes \rho_E$ is the ideal secret state (describing two identical key strings completely decoupled from Eve). 

\section{Epsilon Security in QKD}
The epsilon security $\varepsilon$ can be decomposed into individual parameters, found at various stages throughout the protocol, each associated with an imperfection. Typically, five components have been distinguished:
\begin{itemize}
    \item \textbf{Parameter estimation (PE) error} $\varepsilon_\text{PE}$. This is the probability that the estimated channel parameters do not belong in the marked out confidence region, laid out by the worst-case scenario estimators.
    \item \textbf{Entropy estimation error} $\varepsilon_\text{ent}$. It is associated with the impact of finite samples on the entropy estimation of key generation sequences. This is only present in CV-QKD protocols.
    \item \textbf{Correctness} $\varepsilon_\text{cor}$. It represents the hash collision probability of a family of universal hash functions used during the verification stage of error correction (EC). It is related to the probability of having different key strings after EC.
    \item \textbf{Smoothing error} $\varepsilon_\text{s}$. It quantifies how close the smoothed key distribution is allowed to be to the true one. In other words, it bounds the probability that the true state lies outside the neighborhood of size $\varepsilon_\text{s}$ used for smoothing in the security analysis.
    \item \textbf{Hashing error} $\varepsilon_\text{h}$. This indicates the collision probability of the universal hash function used at the privacy amplification (PA) stage.
\end{itemize}
Note that the latter two parameters are included in the secrecy parameter $\varepsilon_\text{sec}$, which characterizes the overall probability of failure associated with PA, i.e.,
\begin{equation}
    \varepsilon_\text{sec} = \varepsilon_\text{s} + \varepsilon_\text{h}.
    \label{eq:e_sec}
\end{equation}
More specifically, the $\varepsilon$-secrecy parameter bounds the trace distance of the final state (after PA) from the ideal state, where Eve's is completely decoupled.

In CV-QKD protocols, the total epsilon security can be written as follows~\cite{ImprovedRates}
\begin{equation}
    \varepsilon = 3 \varepsilon_\text{PE} + \varepsilon_\text{cor} + \varepsilon_\text{sec},
    \label{eq:cvqkd}
\end{equation}
where the factor $3$ before $\varepsilon_\text{PE}$ is due to the estimation of two different channel parameters, i.e. transmissivity and excess noise, plus to simplify the optimization process, we take the entropy estimation penalty as $\varepsilon_\text{ent} = \varepsilon_\text{PE}$. For the single-photon BB84, the epsilon security is 
\begin{equation}
    \varepsilon = \varepsilon_\text{PE} + \varepsilon_\text{cor} + \varepsilon_\text{sec}.
    \label{eq:bb84_1_photon}
\end{equation} 

\section{Composable Key Rates}
In this section, the review the composable secret key rate for both CV- and DV-QKD. We are interested in the conditional key rate, that is, the rate from a single block of a QKD session, assuming successful EC. The analysis can be easily extended to the unconditional key rate, where we average the performance by considering the probability of success of EC. 

\subsection{CV-QKD Composable Key Rate}
We consider CV-QKD with Gaussian-modulated coherent states (with variance $\mu_{\mathrm{sig}}$) and homodyne detection, with efficiency $\eta$ and electronic noise $v_{\mathrm{el}}$. In our setting, the quantum channel between Alice and Bob is implemented by a fiber link. For a fiber of length $L$ and attenuation $A$, the transmissivity is given by
\begin{equation}
    T = 10 ^ {-\frac{AL}{10}}.
    \label{eq:transmissivity}
\end{equation}
To define the composable key rate $R$, we start with the asymptotic key rate as \cite{ImprovedRates}
\begin{equation}
R_{\infty} = \beta I(\mathbf{x}:\mathbf{y}) - \chi(E:\mathbf{y}),
\label{cvqkd_asymptotic}
\end{equation}
where $\beta$ is the reconciliation efficiency (heuristically set to values $\le 1$), $I(\mathbf{x}:\mathbf{y})$ is Alice and Bob's mutual information, and $\chi(E:\mathbf{y})$ is Eve's Holevo information on Bob's variable.
With the introduction of estimators for the transmissivity $T\rightarrow\hat{T}$ and the excess noise $\xi\rightarrow\hat{\xi}$, obtained during the parameter estimation stage, the rate shown in Eq.~\eqref{cvqkd_asymptotic} becomes
\begin{equation}
    R_\text{PE}= \beta I(\widehat{T},\widehat{\xi})-\chi(T_{\mathrm{wc}},\xi_{\mathrm{wc}}),
    \label{eq:cvqkd_pe_rate}
\end{equation}
where $T_\text{wc}$ and $\xi_\text{wc}$ denote the worst-case scenario estimators, provided by
\begin{align}
    T_\text{wc} = \hat{T} - w \sigma_{\hat{T}},~
    \xi_\text{wc} = \hat{\xi} - w \sigma_{\hat{\xi}}.
\end{align}
Here, $\sigma$ stands for the variance of the respective channel parameter and parameter $w$ is related to the parameter estimation error by
\begin{equation}
    w = \sqrt{2\ln\left(\dfrac{1}{\varepsilon_\text{PE}}\right)}.
\end{equation}
Given that $m$ states are sacrificed for PE, i.e.,
\begin{equation}
    m = r_\text{PE}N
\end{equation}
where $r_\text{PE}$ is the sacrificed state fraction for PE, only $n = N - m$ states will be used for key generation. This means that a factor $n/N$ will multiply the key rate. The composable key rate is given by
\begin{equation}
    R = \frac{1}{N} (n R_{\mathrm{PE}}  - F),
    \label{eq:cvqkd_rate}
\end{equation}
where $F$ accounts for finite-size terms~\footnote{This rate can be written under the assumptions of $\varepsilon_{\mathrm{s}}=\varepsilon_{\mathrm{h}}=\varepsilon_{\mathrm{sec}}/2$ and $\varepsilon_{\mathrm{PE}}=\varepsilon_{\mathrm{ent}}$.}
\begin{multline}
    F= (\sqrt{n} \log_2n)\sqrt{2\ln\left(\dfrac{2}{\varepsilon_\text{PE}}\right)} + 4 \sqrt{n}  \log_{2}  \left(\sqrt{2^{D}}+2\right) \\ \times \sqrt{\log_{2}\left(\dfrac{8}{\varepsilon_{\text{sec}}^{2}}\right)} - \log_2\left(\dfrac{\varepsilon^{2}_{\text{sec}}\varepsilon
    _{\text{cor}}}{2} \right),
    \label{eq:cvqkd_finite}
\end{multline}
with $D$ being the discretization parameter, i.e. the number of bins used to map continuous quadrature outcomes to discrete values for key extraction. For a source with a repetition rate $\text{clk}$, measured in uses/sec, we can write a rate in bits/sec as
\begin{equation}
    R \rightarrow \text{clk} R.
    \label{eq:clk}
\end{equation}

\subsection{DV-QKD Composable Key Rate}
In the single-photon BB84 protocol, the composable key rate in uses/sec is given by
\begin{equation}
    R = \kappa r.
    \label{eq: dvqkd_rate}
\end{equation}
To begin with, the transmissivity in Eq.~\eqref{eq:transmissivity} combines with the detector efficiency $\eta$ to give the overall efficiency
\begin{equation}
    \eta_\text{tot} = \eta T.
    \label{eq:total_efficiency}
\end{equation}
In Eq.~\eqref{eq: dvqkd_rate}, $\kappa$ stands for the raw key rate as
\begin{equation}
    \kappa = (1-r_{\text{PE}}) p_{\text{sift}} Q_1,
\end{equation}
where $r_\text{PE}$ is the sacrificed state ratio for parameter estimation, $p_\text{sift}$ is the sifting probability, calculated from the probability of the $X$-basis, $p_X$, as
\begin{equation}
    p_\text{sift} = p_X^2 + (1 - p_X)^2,
\end{equation}
and $Q_1$ is Bob's detection probability, given by
\begin{equation}
    Q_1 = \eta_\text{tot} + (1 - \eta_\text{tot})p_\text{dc}
\end{equation}
with $p_\text{dc}$ denoting the dark-count probability. 

Assuming a block of size $N=n+m$, where $m$ points are for PE and $n$ are used for key generation, the secret fraction $r$ is given by
\begin{align}
    r&=1-h(\tilde{E})-f_\text{EC}h(\widehat{E}) \notag \\ &+\frac{1+\log_2 (\varepsilon_\text{cor} \varepsilon^2_{\text{h}})}{n}-\frac{\Delta_{\text{AEP}}(\varepsilon_\text{s})}{\sqrt{n}}.
\end{align}
Here the asymptotic equipartition property (APE) term reads
\begin{equation}
    \Delta_{\text{AEP}}(\varepsilon_\text{s}) = 7\sqrt{\log_2 \left(\dfrac{2}{\varepsilon_\text{s}}\right)},\label{AEP1}
\end{equation}
$f_\text{EC} > 1$ is the reconciliation efficiency, and, finally, $\hat{E}$ and $\tilde{E}$ stand for the estimated and worst-case QBER respectively, with
\begin{equation}    \tilde{E}=\widehat{E}+\sqrt{\dfrac{2}{m}\ln\left (\dfrac{m+1}{\varepsilon_\text{PE}}\right )}.
    \label{eq:dvqkd_worst_qber}
\end{equation}
To convert $R$ into a rate in bits per second, both the repetition rate $\text{clk}$ and the detector dead time $t_{\text{dt}}$ must be taken into account, as 
\begin{equation}
    R \rightarrow c_\text{dt} \text{clk} R,
    \label{eq:clk_dvqkd}
\end{equation}
where $c_\text{dt}$ is a reduction factor accounting for the detector dead time, given by
\begin{equation}
    c_{\text{dt}} = \frac{1}{1 + Q_1 t_{\text{dt}}\text{clk}}.
    \label{eq:dead_time_rate}
\end{equation}

\section{The Continuous Genetic Algorithm}
The Continuous Genetic Algorithm (CGA) \cite{CGA1,CGA2,CGA3} is an evolutionary optimization technique inspired by natural selection. It generalizes the traditional genetic algorithm by allowing parameters to take continuous values, making it well-suited for optimal control problems.

\subsection{General Description}
In this framework, a chromosome is a candidate solution represented as a vector of continuous parameters. The quality of each chromosome is measured by a fitness function, which encodes the objective of the optimization. The algorithm evolves a population of chromosomes over successive generations through the following steps:
\begin{itemize}
    \item \textbf{Initialization}: Randomly generate an initial population of candidate solutions.
    \item \textbf{Selection}: Evaluate the fitness of each chromosome and retain the best-performing ones.
    \item \textbf{Pairing}: Select pairs of parent chromosomes probabilistically according to fitness, ensuring fitter candidates reproduce more often.
    \item \textbf{Crossover (Mating)}: Combine parent parameters to generate offspring by forming weighted mixtures of the parent values. In the CGA, unlike in discrete genetic algorithms, these weights are typically chosen randomly from the interval $[0,1]$. This allows the offspring to explore not only the parameter values present in the parents, but also any continuous value between them. This way, the search space is expanded.
    \item \textbf{Mutation}: Randomly replace some parameter values with new random values to maintain diversity and avoid local optima. 
    \item \textbf{Elitism}: The best chromosome is preserved unchanged to guarantee non-decreasing maximum fitness across generations.
\end{itemize}
This iterative process continues until convergence or until a satisfactory fitness level is reached. As the CGA is inherently stochastic, it cannot guarantee the identification of the global optimal value in a finite amount of time. However, it reliably improves candidate solutions, as generations progress.

\subsection{Algorithm Formulation}
Depending on the protocol of interest, the security parameters 
$\varepsilon_{\mathrm{PE}}$, $\varepsilon_{\mathrm{cor}}$ and $\varepsilon_{\mathrm{sec}}$ must satisfy a total epsilon security constraint. For CV-QKD, this constraint is shown in Eq.~\eqref{eq:cvqkd}, while for DV-QKD it is given in Eq.~\eqref{eq:bb84_1_photon}.
In our setting, only the $\varepsilon_{\mathrm{PE}}$ and $\varepsilon_{\mathrm{cor}}$ components
are treated as optimization variables. The $\varepsilon_{\mathrm{sec}}$ component is reconstructed from the input $\varepsilon$ plus the two security components. 
It is important to emphasize that the choice of optimization variables is irrelevant to the optimization problem, because the security parameters are constrained by a single linear condition and any two of them uniquely determine the third. Therefore, optimizing over $\varepsilon_{\mathrm{PE}}$ and $\varepsilon_{\mathrm{cor}}$ 
while solving for $\varepsilon_{\mathrm{sec}}$ is fully equivalent to optimizing over all 
three variables, subject to the same constraint. 

In practice, the genetic algorithm operates in a normalized search space. A chromosome is described by a vector $\mathbf{C}$ of two normalized components (the `genes') $p_1$ and $p_2$, i.e.,
\[
\mathbf{C} = (p_1, p_2) \in [-1,1]^2.
\]
The genes are mapped to the physical domain by the linear transformation
\begin{equation}
    x_i = \frac{p_i + 1}{2}\,(b_i - a_i) + a_i,
    \label{eq:mapping}
\end{equation}
where $x_i \in [a_i,b_i]$ is the physical parameter associated with the gene $p_i$. In particular, we have $x_1 = \varepsilon_{\text{PE}}$ and $x_2 = \varepsilon_{\text{cor}}$.
In our application, each of these physical parameters falls within the range 
$[a_i, b_i] = [10^{-21},\, \varepsilon)$.

Each chromosome $\mathbf{C}$ is assigned a fitness value, defined as the secret key rate produced by the corresponding QKD model. Formally, the fitness function is
\begin{equation}
f(\mathbf{C}) 
= R(\varepsilon_{\mathrm{PE}},\, \varepsilon_{\mathrm{cor}},\, \varepsilon_{\mathrm{sec}}),\label{fitF}
\end{equation}
where the key rate $R$ is computed on the physical variables $\varepsilon_{\mathrm{PE}}$ and $\varepsilon_{\mathrm{cor}}$, together with the value of $\varepsilon_{\mathrm{sec}}$ resulting from the relevant epsilon security constraint.


More specifically, for each generation with population size $N_{\text{pop}}$, the CGA consists of the following steps:
\begin{enumerate}
    \item  \textbf{Initialization:} We randomly generate each chromosome $\mathbf{C}$, by choosing its genes $p_1$ and $p_2$ from a uniform distribution over $[-1,1]^2$.

    \item \textbf{Evaluation:} For each chromosome, we map the genes into the physical parameters $x_1=\varepsilon_{\mathrm{PE}}$ and $x_2=\varepsilon_{\mathrm{cor}}$ according to Eq.~\eqref{eq:mapping} where the range is determined by the input epsilon security $\varepsilon$. The additional epsilon parameter $\varepsilon_{\mathrm{sec}}$ is built in such a way to satisfy the relevant constraint, expressed by Eq.~\eqref{eq:cvqkd} or Eq.~\eqref{eq:bb84_1_photon}. If this constraint is violated (e.g., this may happen when $\varepsilon_{\mathrm{PE}}$ and $\varepsilon_{\mathrm{cor}}$ are too close to $\varepsilon_{\mathrm{sec}}$), the chromosome is assigned the worst possible fitness value and therefore effectively discarded during selection.

    \item \textbf{Selection:} For each chromosome, we compute the value of the fitness function according to Eq.~\eqref{fitF}. Then, the chromosomes are sorted by order of fitness and 
    the top 
    \begin{equation}
        N_{\text{parents}} = \lfloor N_{\text{pop}} \rho_\text{parent}\rfloor
    \end{equation}
    individuals create the parent pool. Here, $\rho_\text{parent}$ stands for the parent selection rate. Then, the number of parents that survive onto the next generation can be determined by 
    \begin{equation}
        N_{\text{survivors}} = \max \{1, \lfloor N_{\text{parents}} \rho_\text{survival} \rfloor\},
    \end{equation}
    where $\rho_\text{survival}$ is the survival rate.
    \item \textbf{Pairing:} From the parent pool, mother--father pairs 
    $(\mathbf{C}^{(m)},\mathbf{C}^{(f)})$ are drawn using fitness-proportional 
    softmax sampling. A mother chromosome $\mathbf{C}^{(m)}$ is randomly selected from the parent pool according to the softmax distribution
    \begin{equation}
        \Pr(m=j)=\frac{e^{f_j}}{\sum_{k=1}^{N_{\text{parents}}} e^{f_k}},
    \end{equation}
    where $f_j := f(\mathbf{C}^{(j)})$ is the fitness of the $j$th chromosome $\mathbf{C}^{(j)}$. For each chosen mother, the father is sampled from the parent pool according to a conditional softmax distribution 
    obtained by excluding the mother, i.e., 
    \begin{equation}
        \Pr(f=j\mid m)
        = \frac{e^{f_j}}{\sum_{\substack{k=1, k\neq m}}^{N_{\text{parents}}} e^{f_k}},
    \end{equation}
    where $j \neq m$. This ensures that the two parents in each pair are distinct while still favoring 
    higher-fitness individuals. Note that higher-fitness chromosomes are more likely, but not guaranteed, to be selected.
    \item \textbf{Crossover:} Offspring chromosomes are generated in the normalized domain $ [-1,1]$ by forming convex combinations of the parent genes $\mathbf{C}^{(m)}=(p_1^{m},p_2^{m})$ and $\mathbf{C}^{(f)}=(p_1^{f},p_2^{f})$. For each offspring and each gene index $i$,
    \begin{equation}
        p_i =
        \begin{cases}
            \gamma\, p^{(m)}_i + (1-\gamma)\, p^{(f)}_i & \text{with} \Pr=\frac{1}{2},\\[4pt]
            p^{(m)}_i & \text{otherwise},
        \end{cases}
     \end{equation}
    where each $\gamma$ is sampled independently from the uniform distribution on $(0,1)$.
    \item \textbf{Mutation and Elitism:} With rate $\rho_\text{mutation}$, the normalized genes $p_i$ of the newly generated offspring and the surviving parents are perturbed by adding Gaussian noise with mean $0$ and standard deviation $0.2$. The mutated values are then clipped to remain in the range $[-1,1]$, which means that any value larger than $1$ is set to $1$ and any value smaller than $-1$ is set to $-1$~\footnote{Note that, because mutation takes place in the normalized domain, where the constraints in Eqs.~\eqref{eq:cvqkd} or~\eqref{eq:bb84_1_photon} are not enforced, infeasible physical parameters may arise after mapping. Such chromosomes are detected during the evaluation step of the next iteration and effectively discarded by being assigned the worst possible fitness value.}. Mutation is applied to prepare candidate solutions for the next iteration. 
    However, the best chromosome in the population is excluded from mutation and is carried over unchanged to the next generation. At least one chromosome must be preserved.

\end{enumerate}
Steps 2–6 of this process are repeated for a set number of iterations $N_\text{iter}$ to produce an approximate maximizer of the key rate $R$. After $N_\text{iter}$ iterations, the optimization outcome is the best chromosome, for which we provide the value of the fitness (the highest fitness) and also the values of the optimal/near-optimal physical parameters.

\section{Case Studies}
For all simulations, the following optimization parameters were used in the CGA:
\begin{itemize}
    \item Population Size $N_\text{pop} = 200$
    \item Iterations $N_\text{iter} = 300$
    \item Mutation Rate $\rho_\text{mutation} = 0.5$
    \item Parent Rate $\rho_\text{parent} = 0.5$
    \item Survival Rate $\rho_\text{survival} = 1$
\end{itemize}
Our practical observations indicate that these values are sufficient to achieve optimal outcomes. Increasing the number of iterations or the population size yields negligible improvement, while considerably slowing down the simulation.

\subsection{Epsilon optimization in CV-QKD}
\begin{figure*}[htb!]
\begin{subfigure}{0.48\textwidth}
    \centering
    \includegraphics[width=\linewidth]{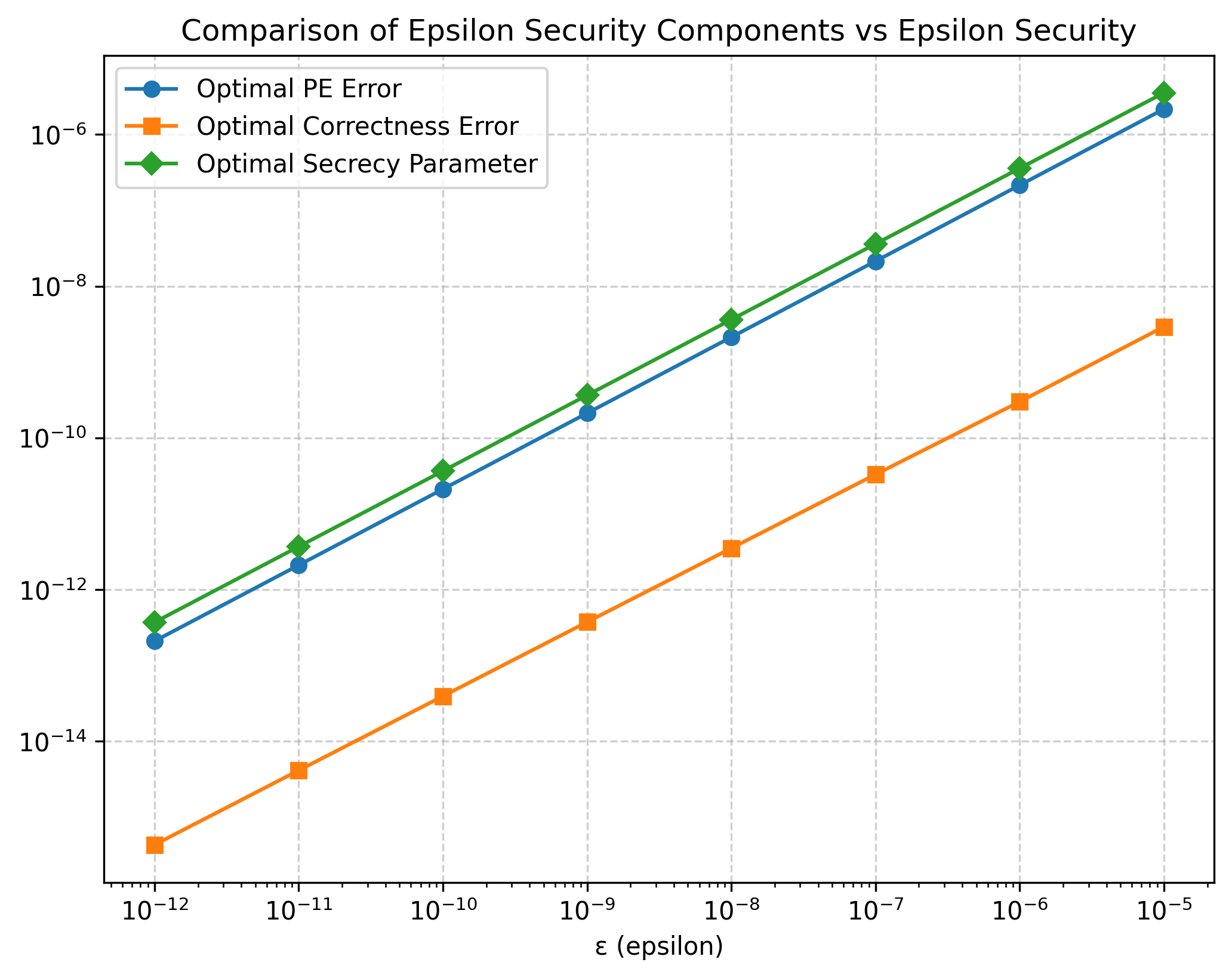}
    \caption{CV-QKD homodyne protocol.}
    \label{fig:cvqkd_params}
\end{subfigure}
\hfill
\begin{subfigure}{0.48\textwidth}
    \centering
    \includegraphics[width=\linewidth]{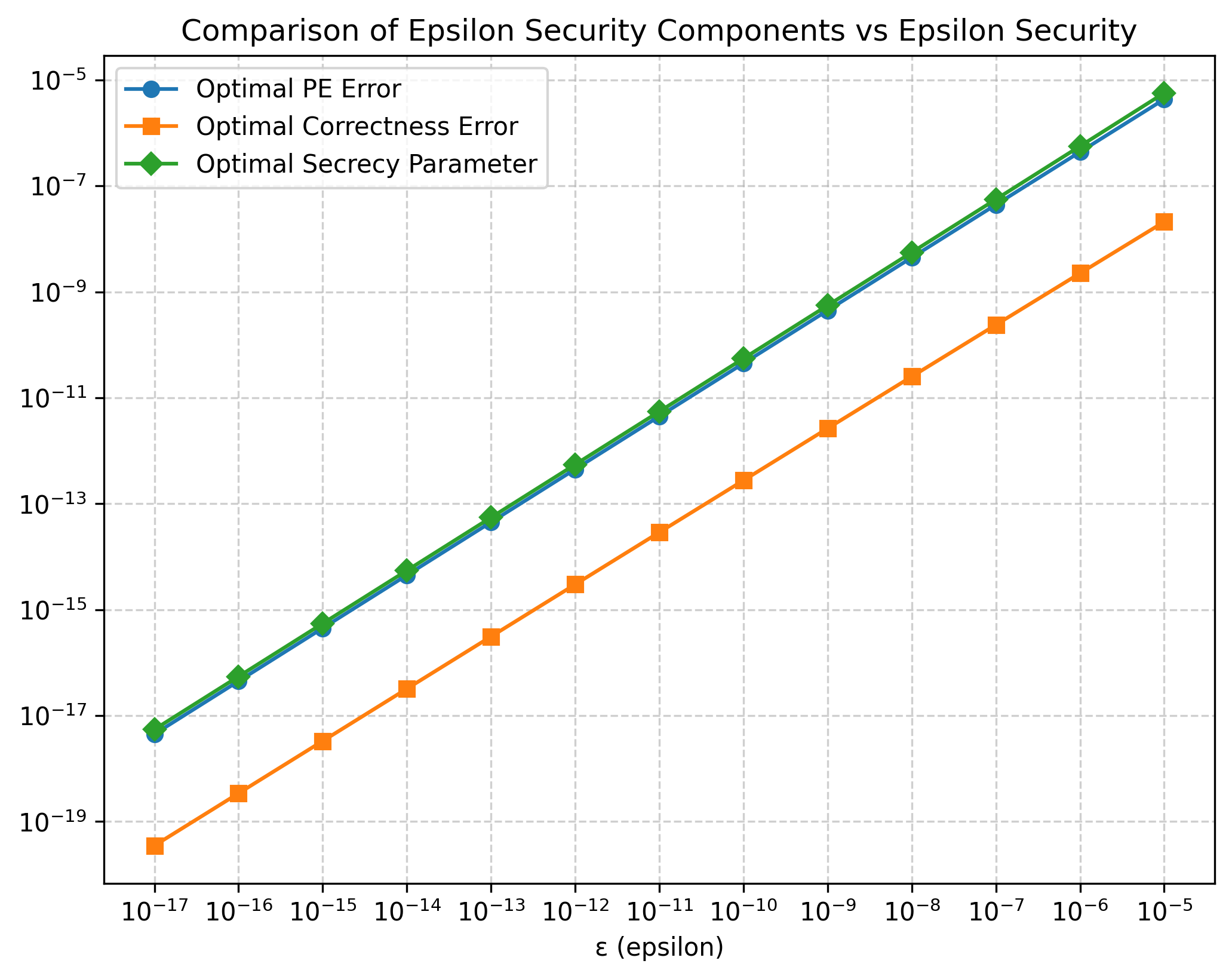}
    \caption{DV-QKD BB84 protocol.}
    \label{fig:dvqkd_params}
\end{subfigure}
\caption{Optimized epsilon security component values vs. epsilon security, for CV-QKD and DV-QKD protocols. For both figures, both axes are plotted on a logarithmic scale.}
\label{fig:e_parameters}
\end{figure*}

For any fixed epsilon security $\varepsilon$, we compare the rate from the optimized parameters $\varepsilon_\text{PE}$, $\varepsilon_\text{cor}$ and $\varepsilon_\text{sec}$ with that from two other cases (one symmetric and the other asymmetric, randomly chosen):
\begin{itemize}
    \item when $\varepsilon_\text{PE} = \varepsilon_\text{cor} = \varepsilon_\text{sec}$ = $\frac{\varepsilon}{5}$ and 
    \item when $\varepsilon_\text{PE} = \frac{\varepsilon}{10}$, $\varepsilon_\text{cor}=\frac{2\varepsilon}{5}$ and $\varepsilon_\text{sec}=\frac{3\varepsilon}{10}$.
\end{itemize}
Both sets satisfy Eq.~\eqref{eq:cvqkd}. The values used for the epsilon security range from $\varepsilon=10^{-12}$ to $10^{-5}$ in steps of one order of magnitude. The resulting optimized epsilon parameters are plotted for these security levels in Fig.~\ref{fig:cvqkd_params}. It is noteworthy that the optimal values of $\varepsilon_{\text{PE}}$ and $\varepsilon_{\text{sec}}$ are extremely close, while $\varepsilon_{\text{cor}}$ is consistently smaller by about two orders of magnitude across all security levels. The rest of the input parameters and their respective values are listed in Table \ref{table:cvqkd}.  

The rate results are displayed in Fig.~\ref{fig:cvqkd}. As seen in the figure, the percentage increase is much higher when the security is also higher, i.e., at lower $\varepsilon$. For very low $\varepsilon$ values, the absolute advantage in the key rate that we obtained by optimizing is minimal. However, as the key rate is closer to zero, even a tiny absolute improvement translates into a large relative percentage gain. By contrast, when the baseline key rate is already high, the change is negligible in percentage terms.

This effect is depicted more clearly in Fig. \ref{fig:cvqkd_zoomed}, where we focus on the interval from $10^{-13}$ to $10^{-12}$ in order to closely examine the behavior in this region. Within this range, the benefits of optimization are evident: it significantly outperforms the case of the randomly-chosen asymmetric parameters and shows a substantial improvement compared to the `division by $5$' approach. Moreover, for $\varepsilon=4 \times 10^{-13}$ and $5 \times 10^{-13}$ optimization yields positive rates, which are unattainable in the other two scenarios. At $\varepsilon=5 \times 10^{-13}$, our optimized approach provides a key rate of about $2$~Mbit/s instead of zero. 


\begin{table}[h!]
\centering
\begin{tabular}{| l | l | l |}
 \hline Parameter & Description & Value   \\ \hline
$L$ & Channel length (km) & $4$  \\
$A$ & Fibre attenuation (dB/km) & $0.2$  \\
$\eta$ & Detector efficiency & $0.85$  \\
$\xi$ & Excess Noise & $0.01$ \\
$v_\text{el}$ & Electronic Noise & $0.1$ \\
$\mu_\text{sig}$ & Signal Variance & $25$ \\
$N$ & Total number of pulses & $4\times10^{5}$   \\
$\beta$ & Reconciliation efficiency ($<1$) & $0.95$  \\
$D$ & Discretization Parameter & $7$ \\
$r_\text{PE}$ & PE ratio & $0.3$ \\
clk & Repetition rate (Hz) & $1 \times 10^9$  \\
\hline
\end{tabular}
\caption{Input parameters for the CV-QKD homodyne protocol.}
\label{table:cvqkd}
\end{table}

\begin{table}[h!]
\centering
\begin{tabular}{| l | l | l |}
 \hline Parameter & Description & Value   \\ \hline
$L$ & Channel length (km) & $100$  \\
$A$ & Fibre attenuation (dB/km) & $0.2$  \\
$\eta$ & Detector efficiency & $0.92$  \\
$E$ & Quantum bit error rate & $\approx 0.05$  \\
$p_X$ & $X$-basis state probability & 0.5\\
$N$ & Total number of pulses & $3\times10^7$   \\
$p_\text{dc}$ & Dark count probability & $1\times 10^{-3}$  \\
$f_\text{EC}$ & Reconciliation efficiency ($>1$) & $1.25$  \\
$r_\text{PE}$ & PE state ratio & $0.25$ \\
clk & Repetition rate (Hz) & $2 \times 10^9$  \\
$t_\text{dt}$ & Detector dead-time (s) & $2 \times10^{-6}$ \\
\hline
\end{tabular}
\caption{Input parameters for the DV-QKD protocol (single-photon BB84).}
\label{table:dvqkd}
\end{table}

\subsection{Epsilon optimization in DV-QKD}
We will compare the key rate produced by the optimal epsilon parameters with the rates of two additional sets of suboptimal parameters that satisfy Eq.~\eqref{eq:bb84_1_photon}, i.e.,
\begin{itemize}
    \item $\varepsilon_\text{PE} = \varepsilon_\text{cor} = \varepsilon_\text{sec} = \frac{\varepsilon}{3}$ and
    \item $\varepsilon_\text{PE}=\frac{5\varepsilon}{99.5}$, $\varepsilon_\text{cor}=\frac{90\varepsilon}{99.5}$, $\varepsilon_\text{sec}=\frac{4.5\varepsilon}{99.5}$.
\end{itemize} 
A comprehensive table of all input parameters that we used can be found in Table \ref{table:dvqkd}. All results are evaluated for a long channel distance of $L = 100$ km. The examined security region spans from $\varepsilon=10^{-17}$ to $10^{-5}$ in decade steps. The values of the optimized epsilon parameters for every security level are shown in Fig. \ref{fig:dvqkd_params}. Similarly to the CV-QKD case, the optimal values of $\varepsilon_{\text{PE}}$ and $\varepsilon_{\text{sec}}$ are nearly identical for every level of $\varepsilon$. In contrast, $\varepsilon_{\text{cor}}$ remains on the order of $10^{-2}$ times the values of $\varepsilon_{\text{PE}}$ and $\varepsilon_{\text{sec}}$ across the entire security range.

\begin{figure*}[htb!]
\begin{subfigure}{0.48\textwidth}
    \centering
    \includegraphics[width=\linewidth]{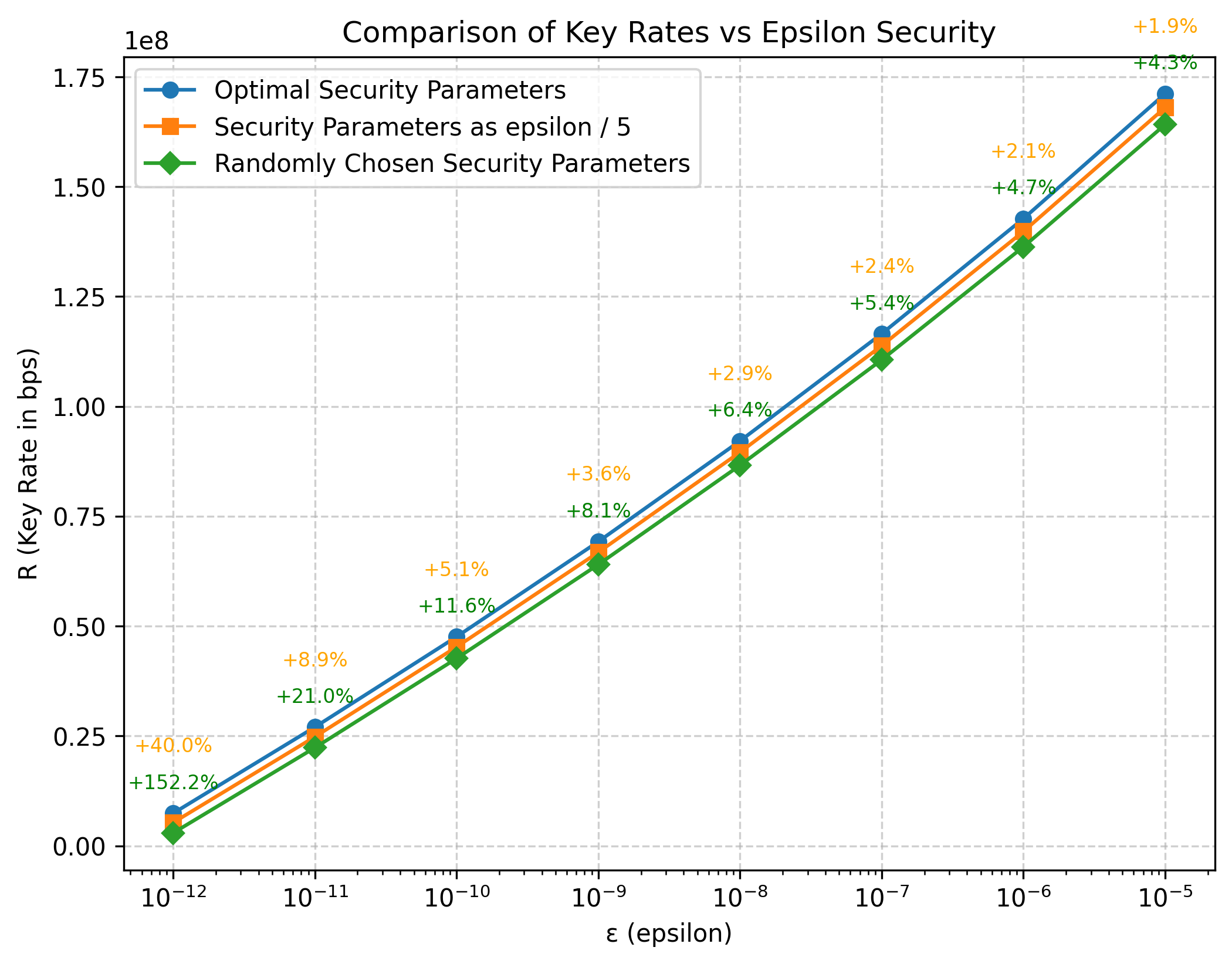}
    \caption{CV-QKD homodyne protocol.}
    \label{fig:cvqkd}
\end{subfigure}
\hfill
\begin{subfigure}{0.46\textwidth}
    \centering
    \includegraphics[width=\linewidth]{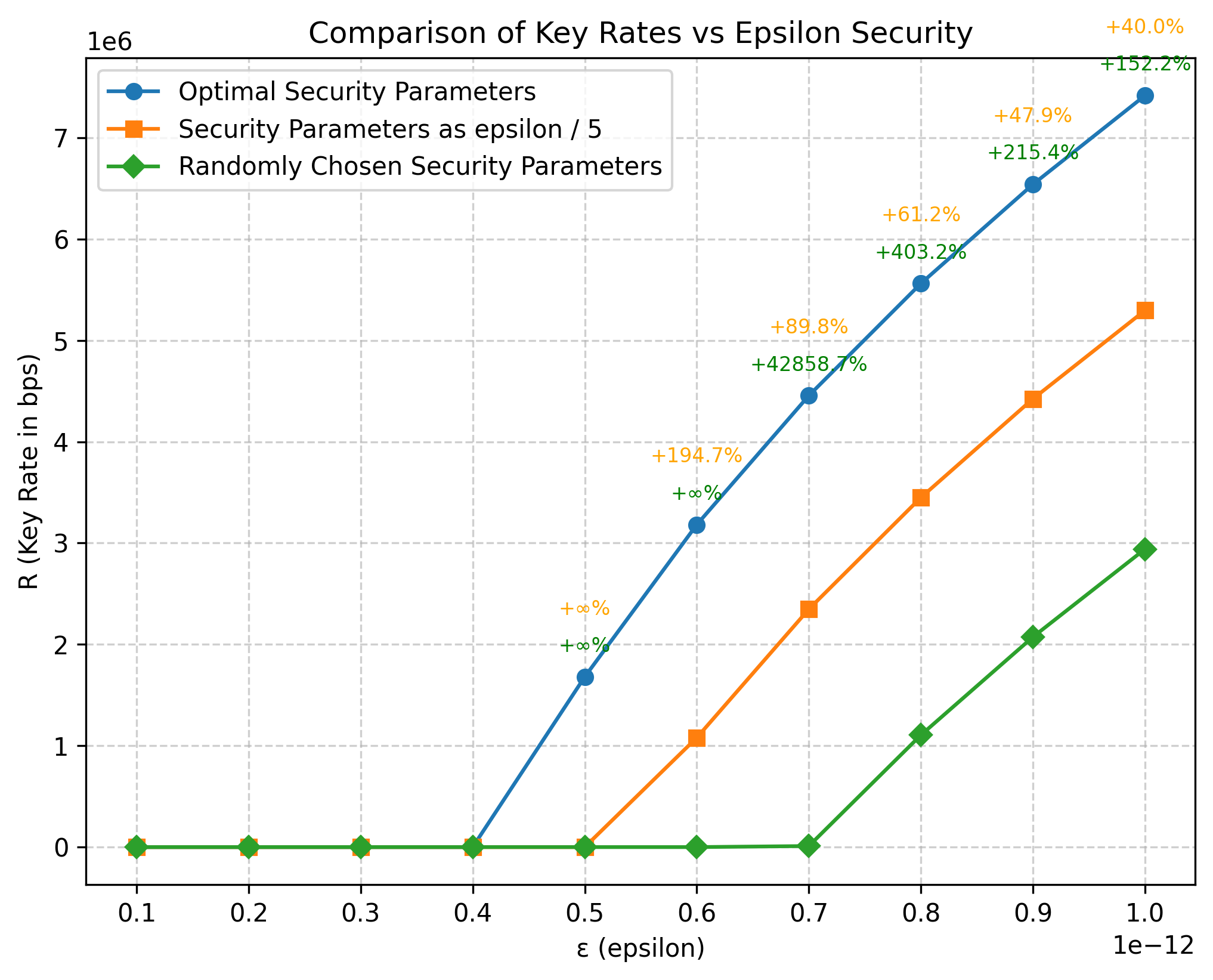}
    \caption{CV-QKD homodyne protocol.}
    \label{fig:cvqkd_zoomed}
\end{subfigure}

\begin{subfigure}{0.48\textwidth}
    \centering
    \includegraphics[width=\linewidth]{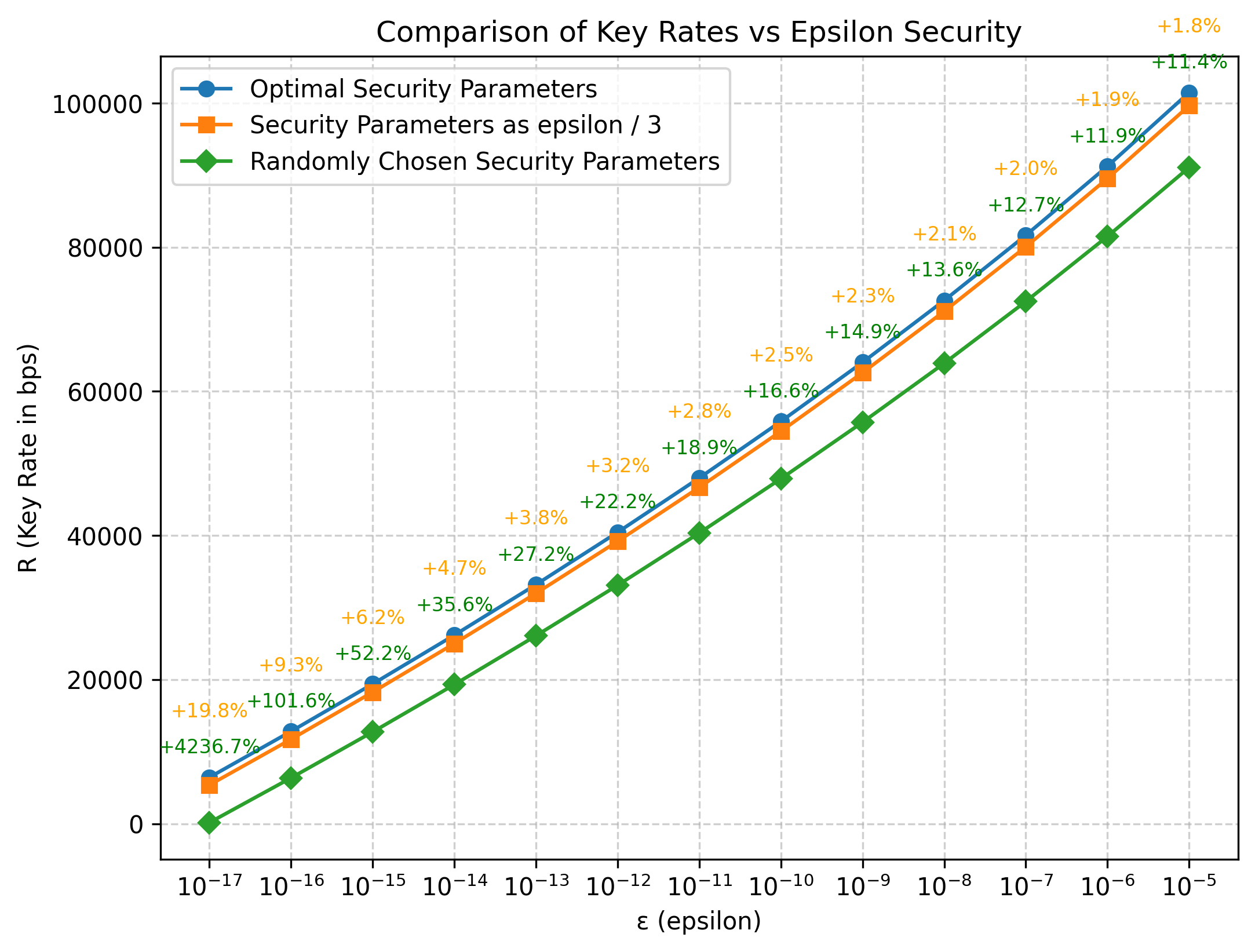}
    \caption{DV-QKD BB84 protocol.}
    \label{fig:bb84}
\end{subfigure}
\hfill
\begin{subfigure}{0.475\textwidth}
    \centering
    \includegraphics[width=\linewidth]{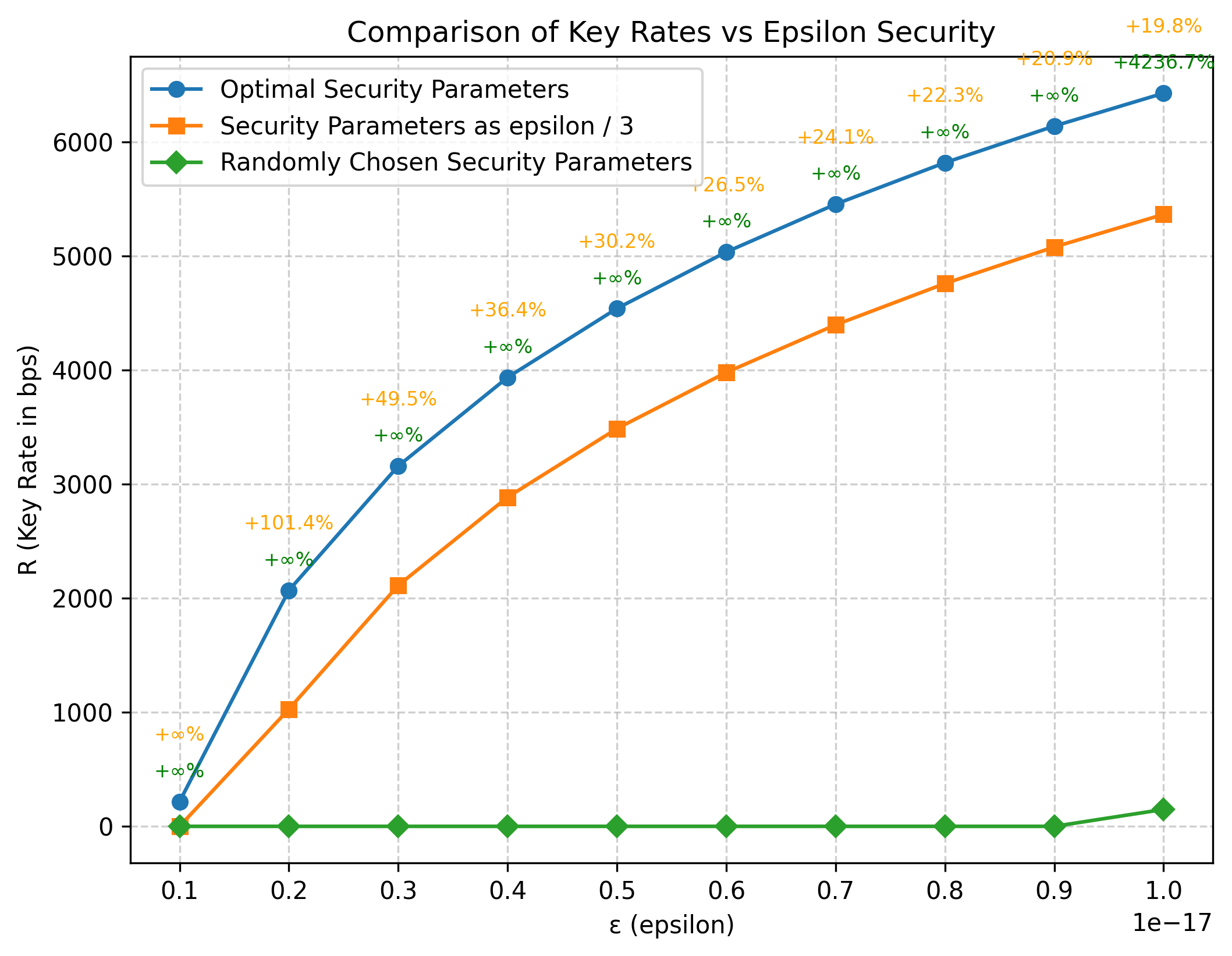}
    \caption{DV-QKD BB84 protocol.}
    \label{fig:bb84_zoomed}
\end{subfigure}
\caption{Composable key rate (bits/sec) vs. epsilon security, for CV-QKD and DV-QKD protocols. For (a) and (c), the x-axis is plotted on a logarithmic scale. For (b) and (d), the x-axis is plotted on a linear scale.}
\label{fig:all_qkd}
\end{figure*}

The composable rate $R$ results are shown in Fig. \ref{fig:bb84}. Again, simply dividing the $\varepsilon$-parameters by a common factor produces results that are close to optimal, though optimization provides a slight improvement. As the key rate approaches zero, optimization achieves increasingly greater percentage gains. This trend becomes particularly evident, when examining the region from $10^{-18}$ to $10^{-17}$, as shown in Fig. \ref{fig:bb84_zoomed}. There, the set of optimized $\varepsilon$-parameters is the only one capable of achieving a positive rate at $\varepsilon=10^{-18}$. For security levels from $\varepsilon=2\times10^{-18}$, the equal values set for the epsilon security constituents can also achieve a positive rate, although it is much smaller. In contrast, the randomly-chosen asymmetric parameters achieve a positive rate only at $\varepsilon=10^{-17}$, showing the importance of optimization in this context.

\section{Conclusion}
We have presented an application of a CGA for the optimization of composable security parameters in QKD protocols. In our formulation, each chromosome encodes three $\varepsilon$-security parameters, specifically the parameter estimation error $\varepsilon_{\text{PE}}$, the correctness error $\varepsilon_{\text{cor}}$ and the secrecy parameter $\varepsilon_{\text{sec}}$ within their admissible domain, and the fitness function is given by the secret key rate $R$. This setting allows us to systematically explore the parameter space and identify configurations that maximize the key rate for fixed overall epsilon security.

For both CV-QKD and DV-QKD BB84 protocols, the same effects were observed by optimizing the $\varepsilon$-parameters. The optimized values of $\varepsilon_\text{PE}$ and $\varepsilon_\text{sec}$ are nearly identical, whereas $\varepsilon_\text{cor}$ remains notably smaller at every considered security level. In addition, while already high key rates at high values of $\varepsilon$-security do not benefit significantly, rates at lower values of $\varepsilon$-security can experience massive improvements. Here $\varepsilon$-parameter optimization can achieve positive rates in regimes where a random or a standard choice fail.

\section*{Acknowledgments}
UKRI supported this work through the Integrated Quantum Networks (IQN) Research Hub (EPSRC, Grant No. EP/Z533208/1).


\end{document}